# Profitability Analysis in Stock Investment Using an LSTM-Based Deep Learning Model

Jaydip Sen
Department of Data Science
Praxis Business School
Kolkata, INDIA
email: jaydip.sen@acm.org

Abhishek Dutta
Department of Data Science
Praxis Business School
Kolkata, INDIA
email: duttaabhishek0601@gmail.com

Sidra Mehtab
Praxis Business School
Kolkata, INDIA
Department of Data Science
email: smehtab@acm.org

*Abstract*— Designing robust systems for precise prediction of future prices of stocks has always been considered a very challenging research problem. Even more challenging is to build a system for constructing an optimum portfolio of stocks based on the forecasted future stock prices. We present a deep learning-based regression model built on a long-and-short-term memory network (LSTM) network that automatically scraps the web and extracts historical stock prices based on a stock's ticker name for a specified pair of start and end dates, and forecasts the future stock prices. We deploy the model on 75 significant stocks chosen from 15 critical sectors of the Indian stock market. For each of the stocks, the model is evaluated for its forecast accuracy. Moreover, the predicted values of the stock prices are used as the basis for investment decisions, and the returns on the investments are computed. Extensive results are presented on the performance of the model. The analysis of the results demonstrates the efficacy and effectiveness of the system and enables us to compare the profitability of the sectors from the point of view of the investors in the stock market.

*Keywords—Stock Price Prediction, Regression, Long and Short-Term Memory Network, Multivariate Time Series, Portfolio Management, Huber Loss, Accuracy Score, MAE.*

## I. INTRODUCTION

Designing robust frameworks for precise prediction of future prices of stocks has always been considered a very challenging research problem. The advocates of the efficient market hypothesis affirm the impossibility of an accurate forecasting of future stock prices. However, propositions demonstrate how complex algorithms and sophisticated and optimally designed predictive models enable one to forecast future stock prices with a high degree of precision. One classical method of predicting stock prices is the decomposition of the stock price time series [1]. Most of the recent works in the literature for forecasting stock prices use approaches based on machine learning and deep learning [2-4]. Bollen et al. contend that emotions profoundly affect an individual's buy or sell decisions [3]. The authors propose a mechanism that computes the public's collective mood from the Twitter feeds and investigate whether the collective moods have any impact on the Dow Jones Industrial Average. The application of convolutional neural networks (CNN) in designing predictive systems for forecasting future stock prices is proposed in some works [4].

Researchers have proposed different approaches to the technical analysis of stocks. Numerous methods have been suggested for dealing with the technical analysis of stock prices. Most of these approaches are based on searching and detecting well-known patterns in the sequence of stock price movement so that the investors may devise profitable strategies upon identifying appropriate patterns in the stock price data. For this purpose, a set of indicators has been recognized for characterizing the stock price patterns.

This paper presents a deep learning-based regression model built on an LSTM network. The system automatically scrapes historical stock prices using its ticker in the NSE and makes a robust prediction of the future values on a daily interval. The model is deployed on 75 stocks chosen from 15 sectors of the Indian stock market, and the return on investment as per the prediction of the model is computed for each stock. The results are analyzed to evaluate the accuracy of forecasting and also for comparing the return on investments for the sectors on which we carry out our study.

The three major contributions of the current work are as follows. First, the work presents a deep learning-based regression model built on an LSTM network that is capable of forecasting future stock prices with a significantly high level of precision. Second, the forecasted output of the LSTM model is used to make investment decisions on 75 different stocks from 15 critical sectors of the stock market of India. The results show the efficacy and effectiveness of the predictive models. Third, the study also enables one to understand the relative profitability of the sectors from the point of view of the investors in the stock market.

We organize the paper as follows. Section II provides a brief discussion on some of the related works. Section III provides a detailed discussion on the data used and the methodology used. Section IV provides the design details of the deep learning regression model built on LSTM. Section V presents extensive results of the performance of the model. Finally, Section VI concludes the paper.

## II. RELATED WORK

The literature on the construction of robust portfolio systems of stocks using sophisticated predictive models is quite rich. Researchers have proposed numerous techniques and approaches for precise forecasting of future movements of stock prices. Broadly, these methods are classified into four types. The works of the first category use different types of regression methods, including the ordinary least square (OLS) regression and its other variants like penalty-based regression, polynomial regression etc. [5-7]. The second category of propositions is based on various econometric methods like autoregressive integrated moving average (ARIMA), cointegration, quartile regression etc. [8-10]. The approaches of the third category use machine learning, deep learning, and reinforcement learning algorithms [11-13]. These approaches are based on predictive models built on complex algorithms and architectures. The works of the fourth category are based on hybrid models built on machine learning and deep learning with inputs of historical stock prices and the sentiments in the news articles on the social web [14-15]. The forecasting performances of these models are found to be the most robust and accurate.

The most difficult challenge in designing a robust and accurate system framework for predicting future stock prices is handling the randomness and the volatility exhibited by the time series. Moreover, the predicted values of stock should be finally used to guide an investor in making a wise investment in the stock market for maximizing the profit out of the investment. Most of the currently existing works in the literature either have not considered this objective or have done it in a somewhat ad-hoc manner. In the current work, we utilize the power of a deep learning model in learning the features of a stock price time series and then making a prediction of its future values with a high level of precision. The predicted stock price is used as a guide for investment in the top five stocks in the fifteen important sectors of the Indian economy. The expected buy and sell profit for each of the stocks is computed, and an overall profit in investment for all the sectors is computed. The work, therefore, shows the relative profitability of the sectors in addition to demonstrating the effectiveness and efficacy of the predictive model.

III. DATA AND METHODOLOGY

As we have mentioned earlier, our goal is to design a predictive model based on LSTM networks and use the predicted values as a guide to investments in various stocks of different sectors of the Indian economy. The profit earned from the investment will provide us an idea about the robustness and accuracy of the models and will also give us an insight into the comparative idea of the profitability of those sectors. We use the Python programming language for developing our proposed system. The proposed system work on the interaction among five functions. The functions are as follows: (1) *load_data*, (2) *create_model*, (3) *comb_df* (4) *plot_graph*, and (5) *predict*. In the following, we describe these in detail.

(1) **load_data**: This function scrapes the data from the web and carries out all preprocessing before the data are used for building the model. The function uses the following parameters: (i) *ticker*, (ii) *no_steps*, (iii) *scale*, (iv) *shuffle*, (v) *forward_step*, (vi) *split_by_date*, (vii) *test_size*, and (viii) *variables_col*. We discuss the role of each parameter in the following.

The parameter *ticker* refers to the ticker of the stock that we want to load. The *yahoo_fin* API in Python is used for extracting the historical records of the stock price data from during a start and an end date mentioned in the program. Our proposed model is constructed using the historical stock price records of the National Stock Exchange (NSE) of India. For example, for the Bajaj Auto stock, we use the ticker string, "BAJAJ-AUTO.NS". Every stock listed in the NSE has a unique ticker string, which is used for extracting the data from the web automatically using the *yahoo_fin* API.

The parameter, *no_steps*, refers to the number of records used in one round of prediction. The default value of this parameter is set to 50, so that we need to feed in our model a sequence of past 50 records to predict the next value in the series. In other words, 50 past records are used to predict the next day's stock price. These values are tunable, however, and can be changed based on the requirement.

The parameter, *scale*, is a Boolean variable that indicates whether to scale the values of the stock prices in the range 0 to 1. By default, this parameter is set to 'True' so that the input values are scaled in the interval [0, 1]. The *MinMaxScalar* function of the *sklearn* module of Python is used for scaling the columns (i.e., the variables). The scaling is needed since the neural network converges faster when the variables are scaled, and the variables with higher values do not get any opportunity to undesirably dominate the model.

The parameter *shuffle* is a Boolean variable, which, if set to *True*, will shuffle the records so that the model does not learn from the sequence of the records in the dataset. The default value of the parameter is *True*.

The *forward_step* parameter refers to the future lookup step to predict. The default value is 1, which implies that the stock price for the next day is predicted.

The parameter *split_by_date* is a Boolean variable whose value determines whether the records are split into the training and test sets by the *date attribute* of the stock price records. When set to *True*, the training and the test datasets will be created based on the date attribute of the records. On the other hand, if it is set to *False*, the training and the test sets will be created by *randomly* choosing the records. The default value of the parameter is set to *True*.

The parameter *test_size* is a *float* variable that indicates the size of the test dataset as a fraction of the size of the total dataset size. Hence, if test_size has a value of 0.2, it indicates that 20% of the total number of records are allocated to the test dataset, and accordingly, the training dataset consists of 80% of the total number of records.

The *variables_col* parameter refers to the list of features of the stock price records used in the model. By default, all the features captured by the *yahoo_fin* API are used in this list. The features are: *adjusted close* (*adj_close*), *open*, *high*, *low*, *close*, and *volume*.

In summary, the load_data function extracts the stock records using the *stock_info.get_data* function in the *yahoo_fin* API. It then adds the date column from the index if it does not exist in the original data. If the *scale* argument is set as True, the function scales all the attributes of the stock price in the interval [0, 1] using the *MinMaxScalar* class of the *sklearn* module in Python. Further, it adds a new column, *future,* that indicates the target values to predict by shifting the *adj_close* column (the target variable) by the value of the parameter, *forward_lookup*. Finally, the function shuffles and divides the data into the training and the test dataset using the date column and returns the training and the test datasets.

(2) **create_model**: This function constructs the deep learning model. We will describe the architecture of the model in the next section. The function *create_model* uses the following arguments: (i) *sequence_length*, (ii) *no_features*, (iii) *no_units*, (iv) *cell_type*, (v) *no_layers*, (vi) *dropout_rate*, (vii) *loss*, (viii) *optimizer*, (ix) *batch_size*, and (x) *epochs*.

The parameter *sequence_length* refers to the number of past records used by the model for predicting. As mentioned earlier, the default length of the input sequence is 50, implying 50 consecutive daily stock price multivariate data are used as the input. In other words, while the value of the parameter sequence_length can be changed based on our requirement, the default value of the parameter is set to 50.

The parameter *no_features* parameter refers to the number of features present in the input stock price data. We use five features: *open*, *high*, *low*, *volume,* and *adjclose*. Among these, the feature *adjclose* is used as the target variable, while the remaining variables are used as the predictors. Hence, the default value of the argument *no_features* is set to 5.

The argument, *no_units*, refers to the number of nodes in each LSTM layer of the model. We will discuss more about this when we will describe the architecture of the model. While the number of LSTM units can be changed based on the complexity of the problem, we use a default value of 256 for this parameter.

The parameter *cell_type* refers to the type of cells used in the recurrent neural network model. As we create an LSTM model, the *cell_type* is set to LSTM.

The parameter, *no_layers*, refers to the number of LSTM layers used in the model. The default value of the parameter is set to be 2, implying two LSTM layers are used. However, based on the complexity of the problem, *no_layers* can also be changed to any desirable value.

The *drop_out* parameter is used to control the training of the model and regularize it. The use of *dropout* prevents overfitting of the model. We will discuss this point further in the architecture of the model in Section IV. In simple terms, *dropout* refers to the fraction of nodes in the LSTM layer that is put off randomly during the training so that the model does not get an opportunity to learn to minutely from the training data. The higher the value used for dropout more likely the model will not be overfitted. However, if the *dropout rate* is too high, the model may enter into an undesirable state of underfitting. We use a default value of 0.3 for the *drop_out* parameter, which is considered a standard value of this purpose.

The parameter *loss* refers to the function used for evaluating the loss exhibited by the model during the training and the validation phase. The loss function can be of multiple types such as *Huber loss* (HL), *mean absolute error* (MAE), *mean square error* (MSE). We use *Huber loss* as the default value of the *loss* parameter.

The optimizer parameter is set to *Adam*. The other possible optimizers are: *stochastic gradient descent* (SGD), *RMSProp*, *Adadelta*, *Adagrad*, *Adamax*, *Nadam*, and *Ftrl*.

The parameter *batchsize* refers to the number of records (i.e., the data points) used in one iteration of the training. An optimum value of *batchsize* minimizes the execution time of an iteration while keep the *batchsize* as large as possible. We find the optimum value of batchsize as 64 using a *gridsearch* method. The optimum value of the *batchsize* is set to 64.

Finally, the parameter, *epochs*, indicates the number of times the *learning algorithm* of the model executes over the entire set of records in the training data. Again, using a *gridsearch* method, we find the optimum value of *epochs* as 100. Hence, the default value of the parameter is set to 100.

In summary, the function *create_model* constructs an LSTM model with an input of sequential data of 50 records and predicts the next value in the sequence using a *solitary node* in its output layer. The learns from the five features in the input data using 256 nodes in the two LSTM layers. It is trained over 100 epochs with a batch size of 64 using the Adam as the optimizer. The Huber loss function is used for evaluating the training and validation process of the model.

(3) *comb_df*: This function takes two parameters, *model* and *data*. The first parameter is the *model* returned by the *create_model* function, while the second parameter is the *data* returned by the *load_data* function. The function *comb_df* constructs a *Pandas Dataframe* object in which it packs the predicted values along with the actual values of stock price. Furthermore, the function performs some important computations using the determination of buy and sell profits from the transactions of a stock. We explain these in the following. The function receives the predicted values of the future stock prices from the return of the function *predict*. We describe the *predict* function later in this section.

If the predicted future price (i.e., the *predicted adj_close* for the next day) exceeds the *current price*, then the function computes the *difference between the actual future price and the current price to compute the buy profit*. In other words, if the model predicts a rise in the price the next day, the investors are advised to buy the stock on the current day. The buy profit is, however, computed based on the difference between the actual price of the stock on the next day and the price of the current day. The stock price here refers to the *adj_close* value, which is the target variable in our study. After computing the buy profits on all possible days in the test cases, the function computes the *total buy profit* by summing them up.

On the other hand, if the predicted future price is lower than the current price, then to compute the sell profit, the function calculates the *difference between the current price and the actual price on the next day*. The function computes the *total sell profit* by summing up all sell profits on the test data points.

The function computes the *total profit* by adding up the *total buy profit* with the *total sell profit*.

Since the number of data samples in the test dataset may not be equal for all stock, we also introduce into this function another task – computation of the values of profit per trade. This value is computed as the *ratio of the total profit to the total number of data points in the test dataset*.

Additionally, the function also computes a *mean absolute score* (MAS) and an *accuracy score* (AS). MAS depicts an average value of error in prediction. A value of 15 indicates that, on average, the predicted values deviate from the actual values by a unit of 15. AS, on the other hand, expresses the fraction of test cases in which the buy or sell profit is positive. If AS for a stock is 0.98, it implies that 98% of the trades on the test data points yielded profit from either sell or buy transactions.

It is important to note that in computing the buy and sell profits, it is assumed that the investor will trade on all the days in the test dataset. If the forecasted price of the following day exceeds the price of the current day, then the investor will buy the stock; else, there will be a sell. The computations are shown on the basis of buy/sell of one stock only. However, if *X* number of shares of a stock are transacted, the buy and sell profits will just get multiplied by the number of shares transacted. In summary, the *comb_df* function uses two arguments- one is the *model* created by the *create_model* function, and the other is the *data dictionary* created by the *load_data* function. It returns a *Dataframe* containing the *features* of the input data samples together with the *actual* and the *forecasted* prices of the stock of the test dataset points.

(4) *plot_graph:* This function takes the Dataframe returned by the function *comb_df*, and plots the actual and the predicted prices (i.e., the *adj_close* values) on the same graph using the *pyplot* function defined in the *matplotlib* module of Python.

(5) *predict:* This function receives two parameters – model and data. The former parameter is the return of the create_model function, while the latter is the return of the load_data function. The function computes the predicted values for all the data points in the test dataset and returns a Dataframe containing the predicted values of the stock prices.

## IV. THE MODEL ARCHITECTURE

In the previous section, we describe in detail the five functions constituting our predictive framework, including the function responsible for data extraction and preprocessing. In this section, we will diss the architecture and further details of the model design. Before discussing the details of the model architecture, we discuss the working principles of LSTM networks and their suitability in handling sequential data such as time series of historical stock price data very briefly.

LSTM is an adaptation of a *recurrent neural network* (RNN) that is capable of interpreting and then forecasting sequential data like time series and text [12]. The networks are capable of maintaining their state information in their designated memory cells which are called *gates*. The state information stored in the memory cells is aggregated with the past information available at the *forget gates*. The *input gates* receive the currently available information, and using the information at the *forget gates* and the *input gates*, the network computes the predicted value for the next time slot and makes it available at the *output gates* [12].

Fig. 1 shows the design of the proposed LSTM model. The model uses daily stock records of the past 50 days as its input, and each record has five features. This is indicated by the input data shape of (50, 5). The input data is passed on the first LSTM layer consisting of 256 nodes. Hence, the output of the LSTM layer has a shape of (50, 256). This indicates that each of the 256 LSTM nodes processes the 50 records from the input and extracts a total of 256 features from each record. The first LSTM layer is followed by a dropout layer that randomly puts off 30% of the nodes at the first LSTM layer in order to prevent any overfitting of the model. A second LSTM layer follows that receives the output from the first LSTM layer and further extracts 256 features from the data. The second LSTM layer is also controlled by a dropout layer with a 30% dropout rate. The output of the second LSTM layer is passed on finally to a dense layer having 256 nodes at its input and one node at the output. The single node at the output of the dense layer produces the final output of the model as the forecasted value of adjusted close for the next day. The training and validation of the model is carried out over 100 *epochs* using a *batch-size* of 64. The activation function used at all layers in the model is the *rectified linear unit* (ReLU) except for the final output layer, in which the *sigmoid activation function* is used. The *Huber loss* function is used to compute the loss, and the *mean absolute error* (MAE) is used for computing the error. The performance results of the model presented in Section IV are achieved using these parameters. However, alternative choices can be made, and the performance of the model can be studied further. The values of the parameters, *epoch* and *batchsize*, are, however, determined using a *gridsearch* method and should not be changed.

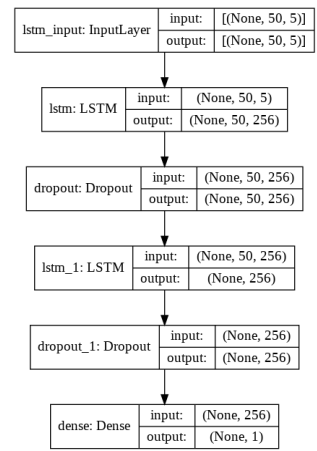

Fig. 1. The architecture of the proposed LSTM model

We use the *Huber loss* function as it effectively acts as a combination of the *mean squared error* (MSE) and the *mean absolute error* (MAE) [16]. The Huber loss function is quadratic when the error is smaller than a threshold value, while it behaves in a linear way when the error is larger than the threshold. When the function is linear, it is less sensitive to the outliers than the MSE. On the other hand, the quadratic part allows it to converge faster and yield more accurate results than the MAE.

## V. EXPERIMENTAL RESULTS

This section presents extensive experimental results on the performance of the predictive model. We apply the model to forecast the adj_close values of stock price for the next day. Based on the forecasted price and the current price of the stock, the predictive model first determines whether a buy or a sell strategy will be recommended to the investor. While the buy profit is computed as the difference between the current price and the actual price of the stock on the next day, the sell profit is determined by calculating the difference between the actual stock price on the next day and the current price of the stock. It is assumed that trading (either buy or sell) is carried out on all the days in the test data. Total profit is computed as the sum of the total buy profit and the total sell profit. Since stock prices widely vary for different companies, we compute the ratio of the total profit to the mean stock price for the data points in the test case. The value of the ratio gives a robust measure of the profitability measure of the investment in the stock. The predictive system also computes the total number of sample points in the test set for each stock and computes the ratio of the total profit to the number of sample points in the test dataset. The value of this ratio is a measure of the profit earned per trade (i.e., buy or sell transaction) for the stock.

We choose fifteen sectors of the Indian economy. These fifteen sectors are: (i) *auto*, (ii) *banking*, (iii) *consumer durables*, (iv) *capital goods*, (v) *fast-moving consumer goods* (FMCG), (vi) *healthcare*, (vii) *information technology* (IT), (viii) *large-cap*, (ix) *metal*, (x) *mid-cap*, (xi) *oil and gas*, (xii) *power*, (xiii) *realty*, (xiv) *small-cap*, and (xv) *telecom*. For each of these sectors, we identify the top five stocks which have the most significant impact on the index of the corresponding sector. The predictive model is deployed to compute the profit in investment in the stocks. Finally, the average of the ratio of the total profit to the mean value of the stock over the test period is computed. This metric is used as a measure of the profit (i.e., the return) of investment for a sector. The models are implemented using Python 3.7.4 on TensorFlow 2.3.0 and Keras 2.4.5 frameworks. The epochs are run on the Google Colab GPU runtime environment. Each epoch approximately took 1 second to execute. In the following, we present the results of all the sectors. It may be noted that the experiments were carried out from March 3 to March 5, 2021. Hence, the stock price on the next day refers to prices from March 4 to March 6.

*Auto sector:* The five significant auto sector stocks listed in NSE are: (i) Maruti Suzuki India (MSU), (ii) Mahindra and Mahindra (MMH), (iii) Tata Motors (TMO), (iv) Bajaj Auto (BAJ), and (v) Hero MotoCorp (HMC). The weights in percent values for these stocks for computing the overall index of the auto sector index in NSE are as follows: MSU-18.88, MMH-15.97, TMO-11.97, BAJ-10.23, HMC-8.66 [17]. For all these stocks, we deploy our proposed predictive framework to compute the following: predicted price on the next day, Huber loss, mean absolute error, accuracy score, total buy profit, total sell profit, total profit, mean stock price over the test period, the ratio of the total profit to the mean stock price, the number of test sample points, the profit per trade computed as the total profit per test sample points. Finally, the average of the ratio of the total profit to the mean stock price is computed for the five stocks to arrive at a measure of the profitability of investment for the overall sector. Table 1 presents the results for the *auto* sector stocks. As an example, Fig. 2 depicts the convergence of the Huber loss in training and validation for the Maruti Suzuki stock.

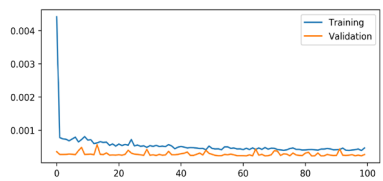

Fig. 2. The Huber loss convergence for the stock data of Maruti Suzuki. The *x*-axis plots the number of epochs, and the *y*- axis, the Huber loss

TABLE I. THE RESULTS OF THE AUTO SECTOR STOCKS

| Metric | BAJ | HMC | MMH | MSU | TMO |
|---|---|---|---|---|---|
| Next day's pred. price | 4189 | 3362 | 843 | 7446 | 333 |
| Huber loss | 0.00012 | 0.00028 | 0.00020 | 0.00023 | 0.00024 |
| Mean absolute error | 109.87 | 155.70 | 16.43 | 256.90 | 17.27 |
| Accuracy score | 0.9846 | 0.9770 | 0.9792 | 0.9837 | 0.9816 |
| Total buy profit | 503415 | 512228 | 181990 | 1166205 | 102313 |
| Total sell profit | 498601 | 509230 | 181077 | 1178960 | 101943 |
| Total profit | 1002016 | 1021458 | 363067 | 2345165 | 204256 |
| Mean stock price | 1307 | 1484 | 270 | 2675 | 162 |
| Total profit/ Mean price | 767 | 688 | 1344 | 877 | 1260 |
| Number of test cases | 911 | 912 | 1252 | 858 | 1253 |
| Profit per trade | 1099.91 | 1120.02 | 289.99 | 2733.29 | 163.01 |
| Avg. profit/mean price | 987 | | | | |

*Banking sector:* The five critical stocks in this sector as per the listing in the NSE and their corresponding weights in percent are: HDFC Bank (HDB)-26.06, ICICI Bank (ICB)-19.56, Axis Bank (AXB)-15.93, State Bank of India (SBI)-13.27, and Kotak Mahindra Bank (KTB)-12.37. Table II presents the results for the *banking* sector stocks. As an example, the plot of the predicted and the actual price for the State Bank of India stock is presented in Fig. 3.

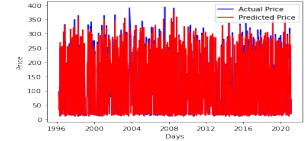

Fig. 3. The actual and the predicted price of the SBI stock over the entire training and test periods – the output of the *plot_graph* function

TABLE II. THE RESULTS OF THE BANKING SECTOR STOCKS

| Metric | AXB | HDB | ICB | KTB | SBI |
|---|---|---|---|---|---|
| Next days' pred. price | 730 | 1573 | 620 | 1874 | 383 |
| Huber loss | 0.00028 | 7.4567 | 0.00023 | 0.00012 | 0.00033 |
| Mean absolute error | 12.47 | 10.78 | 23.14 | 19.73 | 16.24 |
| Accuracy score | 0.9663 | 0.9880 | 0.9780 | 0.9793 | 0.9704 |
| Total buy profit | 136772 | 226446 | 57076 | 241653 | 75752 |
| Total sell profit | 135801 | 228576 | 57450 | 238510 | 74194 |
| Total profit | 272573 | 455022 | 114526 | 480163 | 149946 |
| Mean stock price | 227 | 286 | 178 | 441 | 135 |
| Total profit/ Mean price | 1201 | 1591 | 643 | 1089 | 1111 |
| Number of test cases | 1099 | 1250 | 910 | 964 | 1252 |
| Profit per trade | 248.02 | 364.02 | 125.85 | 498.09 | 119.76 |
| Avg. profit/mean price | 1127 | | | | |

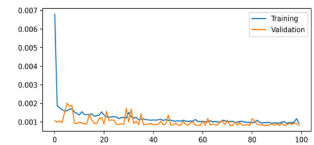

Fig. 4. The training and the validation for the stock data of ABB. The *x*-axis plots the number of epochs, and the *y*- axis, the Huber loss

TABLE III. THE RESULTS OF THE CAPITAL GOODS SECTOR STOCKS

| Metric | ABB | BEL | HVL | LNT | SIM |
|---|---|---|---|---|---|
| Next day's pred. price | 1481 | 135 | 1170 | 1534 | 1844 |
| Huber loss | 0.00080 | 0.00024 | 0.00012 | 0.00030 | 0.00026 |
| Mean absolute error | 43.12 | 5.67 | 11.01 | 52.36 | 40.28 |
| Accuracy score | 0.9671 | 0.9825 | 0.9748 | 0.9737 | 0.9817 |
| Total buy profit | 250798 | 20596 | 113857 | 200786 | 284522 |
| Total sell profit | 251999 | 20824 | 112653 | 200094 | 285985 |
| Total profit | 502797 | 41420 | 226510 | 400880 | 570507 |
| Mean stock price | 729 | 56 | 202 | 614 | 568 |
| Total profit/ Mean price | 690 | 740 | 1121 | 653 | 1004 |
| Number of test cases | 911 | 912 | 913 | 911 | 1095 |
| Profit per trade | 551.71 | 45.42 | 248.09 | 440.04 | 521.01 |
| Avg. profit/mean price | 842 | | | | |

*Capital Goods sector:* The top five stocks in this sector as per the listing in the NSE are: ABB, Bharat Electronics (BEL), Havells India (HVL), Larsen and Toubro (LNT), and Siemens India (SIM). Table III presents the results for the *capital goods* sector stocks. As an example, in Fig. 4, we show the convergence of the Huber loss in the training and validation for the ABB stock.

TABLE IV. THE RESULTS OF THE CONSUMER DURABLE SECTOR STOCKS

| Metric | BAT | HVL | TIT | VOL | WHR |
|---|---|---|---|---|---|
| Next day's pred. price | 1459 | 1170 | 1412 | 1074 | 2385 |
| Huber loss | 0.00015 | 0.00012 | 0.00018 | 0.00013 | 0.00022 |
| Mean absolute error | 27.96 | 11.01 | 13.11 | 10.48 | 41.29 |
| Accuracy score | 0.9846 | 0.9748 | 0.9561 | 0.9683 | 0.8236 |
| Total buy profit | 201396 | 113857 | 212622 | 106242 | 295898 |
| Total sell profit | 200212 | 112653 | 210254 | 103007 | 297410 |
| Total profit | 401608 | 226510 | 422876 | 209249 | 593308 |
| Mean stock price | 398 | 202 | 236 | 207 | 502 |
| Total profit/ Mean price | 1009 | 1121 | 1792 | 1011 | 1182 |
| Number of test cases | 912 | 913 | 1253 | 914 | 907 |
| Profit per trade | 440.36 | 248.09 | 337.49 | 228.94 | 654.14 |
| Avg. profit/mean price | 1123 | | | | |

*Consumer Durable sector:* The top five stocks in this sector as per the listing in the NSE are: Titan Company (TIT), Havells India (HVL), Voltas (VOL), Whirlpool India (WHR), and Bata India (BAT). Table IV presents the results for the *consumer durable goods* sector stocks. As an example, in Fig. 5, we show the plot of the predicted and the actual values of the stock price of Havells India.

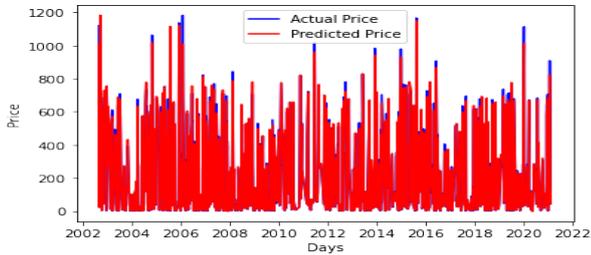

Fig. 5. The actual and the predicted price of the Havells India stock over the entire training and test periods – the output of the *plot_graph* function

*FMCG sector:* The critically important stocks of this sector and their corresponding percentage weights in NSE are: Hindustan Unilever (HUL)-26.80, ITC (ITC)-25.08, Nestle India (NST)-8.09, Britannia Industries (BRT)-6.22, and Dabur India (DBR)-4.46. Table V presents the results for the FMCG sector stocks. The plot of the convergence of the Huber loss in the training and validation for the HUL stock is depicted in Fig. 6.

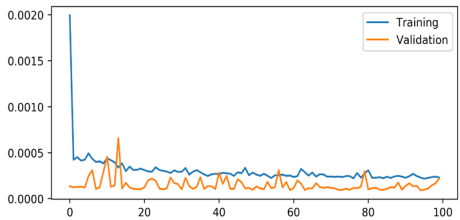

Fig. 6. The training and the validation for the stock data of HUL. The *x*-axis plots the number of epochs, and the *y*- axis, the Huber loss

TABLE V. THE RESULTS OF THE FMCG SECTOR STOCKS

| Metric | BRT | DBR | HUL | ITC | NST |
|---|---|---|---|---|---|
| Next day's pred. price | 3486 | 515 | 2233 | 225 | 16929 |
| Huber loss | 0.00011 | 0.00015 | 0.00009 | 0.00018 | 0.00012 |
| Mean absolute error | 32.92 | 10.06 | 55.77 | 6.27 | 566.77 |
| Accuracy score | 0.9792 | 0.9802 | 0.9656 | 0.9808 | 0.8530 |
| Total buy profit | 574073 | 71087 | 334861 | 63580 | 1895115 |
| Total sell profit | 574696 | 70639 | 336192 | 62254 | 1878462 |
| Total profit | 1148769 | 141726 | 671053 | 125834 | 3773577 |
| Mean stock price | 647 | 152 | 472 | 96 | 3878 |
| Total profit/ Mean price | 1776 | 932 | 1422 | 1311 | 973 |
| Number of test cases | 1249 | 910 | 1250 | 1249 | 905 |
| Profit per trade | 919.75 | 155.74 | 536.84 | 100.75 | 4169.70 |
| Avg. profit/mean price | 1283 | | | | |

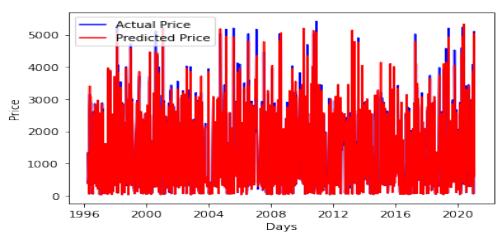

Fig. 7. The actual and the predicted price of the stock of Dr. Reddy's Lab over the entire training and test periods – the output of *plot_graph* function

*Healthcare sector:* The five most critical stocks of listed in the NSE and their corresponding percentage weights are: Sun Pharmaceutical Industries (SNP)-15.99, Dr. Reddy's Laboratories (DRL)-13.38, Divi's Laboratories (DVL)-10.67, Cipla (CPL)-9.96, and Aurobindo Pharma (ARP)-5.99. Table VI presents the results of this sector. Fig. 7 depicts the predicted and the actual price of the stock of Dr. Reddy's Lab.

TABLE VI. THE RESULTS OF THE HEALTHCARE SECTOR STOCKS

| Metric | ARP | CPL | DVL | DRL | SNP |
|---|---|---|---|---|---|
| Next day's pred. price | 842 | 787 | 3458 | 4422 | 601 |
| Huber loss | 0.00023 | 0.00018 | 0.00008 | 0.00017 | 0.00012 |
| Mean absolute error | 13.30 | 14.18 | 36.62 | 85.26 | 10.70 |
| Accuracy score | 0.9759 | 0.9792 | 0.9658 | 0.9856 | 0.9705 |
| Total buy profit | 170208 | 156522 | 285260 | 809723 | 176762 |
| Total sell profit | 168041 | 154150 | 273494 | 801887 | 173076 |
| Total profit | 338249 | 310672 | 558754 | 1611610 | 349838 |
| Mean stock price | 214 | 268 | 633 | 1282 | 235 |
| Total profit/ Mean price | 1581 | 1159 | 883 | 1257 | 1489 |
| Number of test cases | 1246 | 1248 | 877 | 1251 | 1253 |
| Profit per trade | 271.47 | 248.94 | 637.12 | 1288 | 279.20 |
| Avg. profit/mean price | 1274 | | | | |

*IT sector:* The most significant stocks of this sector and with their corresponding percentage weights are: Infosys (IFY)-26.42, Tata Consultancy Services (TCS)-25.92, Wipro (WIP)-9.84, Info Edge India (INE)-9.29, and Tech Mahindra (TEM)-8.89. Table VII presents the results for the IT sector stocks. The plot of the convergence of the Huber loss in the training and validation for the TCS stock is shown in Fig. 8.

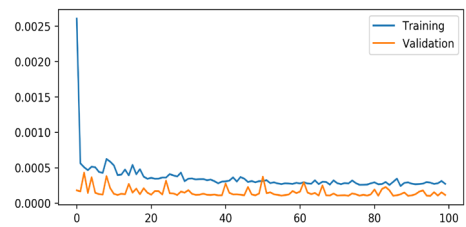

Fig. 8. The training and the validation for the stock data of TCS. The *x*-axis plots the number of epochs, and the *y*- axis, the Huber loss

TABLE VII. THE RESULTS OF THE IT SECTOR STOCKS

| Metric | INE | IFY | TCS | TEM | WIP |
|---|---|---|---|---|---|
| Next day's pred. price | 5090 | 1313 | 3244 | 941 | 460 |
| Huber loss | 0.00011 | 0.00007 | 0.00010 | 0.00021 | 0.00016 |
| Mean absolute error | 133.84 | 9.50 | 117.87 | 59.21 | 5.73 |
| Accuracy score | 0.9811 | 0.9800 | 0.9774 | 0.9743 | 0.9816 |
| Total buy profit | 281802 | 160038 | 287062 | 83785 | 58701 |
| Total sell profit | 279930 | 158469 | 282739 | 81012 | 58560 |
| Total profit | 561732 | 318507 | 569801 | 164797 | 117261 |
| Mean stock price | 822 | 243 | 791 | 366 | 115 |
| Total profit/ Mean price | 683 | 1311 | 720 | 450 | 1020 |
| Number of test cases | 689 | 1250 | 798 | 700 | 1250 |
| Profit per trade | 815.29 | 254.81 | 714.04 | 235.42 | 93.81 |
| Avg. profit/mean price | 837 | | | | |

*Large-Cap sector:* The most significant stocks of this sector in the NSE are: Reliance Industries (RIL), Tata Consultancy Services (TCS), HDFC Bank (HDB), ICICI Bank (ICB), and the Kotak Mahindra Bank (KTB). Table VIII presents the results for the large-cap sector stocks. The plot of the predicted and the actual price for the stock of Reliance Industries is presented in Fig. 9.

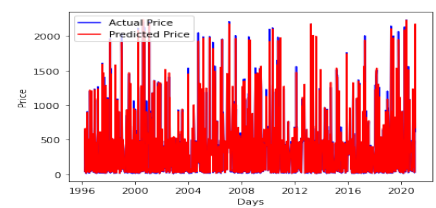

Fig. 9. The actual and the predicted *adjusted close price* of the stock of Reliance Industries over the entire training and test periods

TABLE VIII.  THE RESULTS OF THE LARGE CAP SECTOR STOCKS

| Metric | HDB | ICB | KTB | RIL | TCS |
|---|---|---|---|---|---|
| Next day's pred. price | 1573 | 620 | 1874 | 20558 | 3244 |
| Huber loss | 7.4567 | 0.00023 | 0.00012 | 0.00008 | 0.00010 |
| Mean absolute error | 10.78 | 23.14 | 19.73 | 25.59 | 117.87 |
| Accuracy score | 0.9880 | 0.9780 | 0.9793 | 0.9720 | 0.9774 |
| Total buy profit | 226446 | 57076 | 241653 | 269127 | 287062 |
| Total sell profit | 228576 | 57450 | 238510 | 261771 | 282739 |
| Total profit | 455022 | 114526 | 480163 | 530898 | 569801 |
| Mean stock price | 286 | 178 | 441 | 392 | 791 |
| Total profit/ Mean price | 1591 | 643 | 1089 | 1354 | 720 |
| Number of test cases | 1250 | 910 | 964 | 1251 | 798 |
| Profit per trade | 364.02 | 125.85 | 498.09 | 424.38 | 714.04 |
| Avg. profit/mean price | 1079 | | | | |

*Metal sector:* The five most significant stocks of this sector in the NSE and their corresponding weights are: Tata Steel (TSL)-22.47, Hindalco Industries (HIN)-20.68, JSW Steel (JSW)-15.92, Coal India (CIL)-13.28, and Jindal Steel and Power (JIN)-5.71. Table IX presents the results for the metal sector stocks. Fig. 10 shows the plot of the Huber loss in the training and validation for the Tata Steel stock.

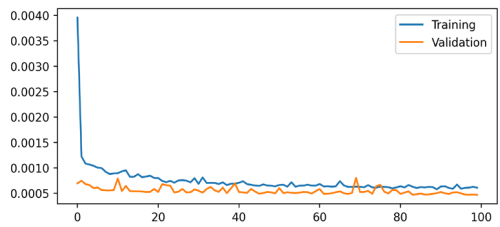

Fig. 10. The training and the validation for the stock data of Tata Steel. The *x*-axis plots the number of epochs, and the *y*- axis, the Huber loss

TABLE IX.  THE RESULTS OF THE METAL SECTOR STOCKS

| Metric | CIL | HIN | JIN | JSW | TSL |
|---|---|---|---|---|---|
| Next day's pred. price | 140 | 360 | 341 | 402 | 723 |
| Huber loss | 0.00107 | 0.00034 | 0.00041 | 0.00025 | 0.00047 |
| Mean absolute error | 111.04 | 32.09 | 14.43 | 13.71 | 35.15 |
| Accuracy score | 0.9677 | 0.9712 | 0.9811 | 0.9804 | 0.9679 |
| Total buy profit | 9091 | 40389 | 112455 | 39809 | 129855 |
| Total sell profit | 9074 | 39947 | 110653 | 40126 | 132074 |
| Total profit | 18165 | 80336 | 223108 | 79935 | 261929 |
| Mean stock price | 198 | 109 | 186 | 107 | 256 |
| Total profit/ Mean price | 92 | 737 | 1199 | 747 | 1023 |
| Number of test cases | 496 | 1252 | 1058 | 869 | 1248 |
| Profit per trade | 36.62 | 64.17 | 210.88 | 91.98 | 209.88 |
| Avg. profit/mean price | 760 | | | | |

*Mid-Cap sector:* The critically important stocks in this sector listed in the NSE are: Adani Enterprise (ADE), Berger Paints (BRP), Biocon (BIO), Info Edge India (INE), and Torrent Pharmaceuticals (TRP). Table X presents the results for the mid-cap sector stocks. The plot of the predicted and the actual price for the stock of Adani Enterprises is presented in Fig. 11.

TABLE X.  THE RESULTS OF THE MID CAP SECTOR STOCKS

| Metric | ADE | BRP | BIO | INE | TRP |
|---|---|---|---|---|---|
| Next day's pred. price | 927 | 706 | 395 | 5090 | 2471 |
| Huber loss | 0.00007 | 0.00009 | 0.00016 | 0.00011 | 0.00014 |
| Mean absolute error | 5.69 | 5.64 | 17.84 | 133.84 | 41.11 |
| Accuracy score | 0.9485 | 0.9792 | 0.9781 | 0.9811 | 0.9606 |
| Total buy profit | 31580 | 70206 | 42962 | 281802 | 338758 |
| Total sell profit | 28117 | 70051 | 42780 | 279930 | 340378 |
| Total profit | 59807 | 140257 | 85742 | 561732 | 679136 |
| Mean stock price | 59 | 117 | 105 | 822 | 626 |
| Total profit/ Mean price | 1014 | 1199 | 817 | 683 | 1085 |
| Number of test cases | 913 | 913 | 821 | 689 | 913 |
| Profit per trade | 65.51 | 153.62 | 104.44 | 815.29 | 744 |
| Avg. profit/mean price | 960 | | | | |

*Oil and Gas Sector:* The stock of significance in this sector and their corresponding percentage weights are: Reliance Industries (RIL)-30.92, Oil and Natural Gas Corporation (ONG)-11.98, Bharat Petroleum Corporation (BPL)-10.68, GAIL India (GAL)-7.76, and Indian Oil Corporation (IOC)-7.38. Table XI presents the results for the metal sector stocks. Fig. 12 shows the plot of the Huber loss in the training and validation for the Oil and Natural Gas Corporation stock.

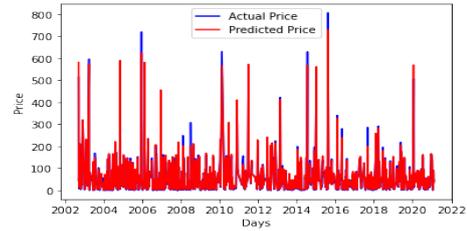

Fig. 11. The actual and the predicted *adjusted close price* of the stock of Adani Enterprise over the entire training and test periods

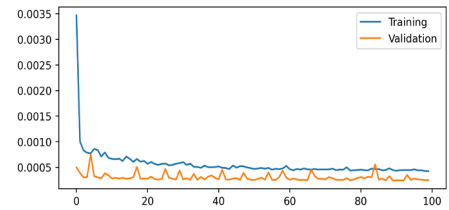

Fig. 12. The training and the validation for the stock data of Oil and Natural Gas Corporation. *x*-axis: the no. of epochs, y-axis: the Huber loss

TABLE XI.  THE RESULTS OF THE OIL AND GAS SECTOR STOCKS

| Metric | BPL | GAL | IOC | ONG | RIL |
|---|---|---|---|---|---|
| Next day's pred. price | 439 | 145 | 98 | 114 | 20558 |
| Huber loss | 0.00021 | 0.00023 | 0.00021 | 0.00024 | 0.00008 |
| Mean absolute error | 9.97 | 6.70 | 5.95 | 7.60 | 25.59 |
| Accuracy score | 0.9672 | 0.9831 | 0.9798 | 0.9832 | 0.9720 |
| Total buy profit | 78775 | 31147 | 27685 | 41974 | 269127 |
| Total sell profit | 79101 | 30990 | 27460 | 41305 | 261771 |
| Total profit | 157876 | 62137 | 55145 | 83279 | 530898 |
| Mean stock price | 109 | 61 | 48 | 83 | 392 |
| Total profit/ Mean price | 1448 | 1019 | 1149 | 1003 | 1354 |
| Number of test cases | 1251 | 1186 | 1236 | 1247 | 1251 |
| Profit per trade | 126.20 | 52.39 | 44.62 | 66.78 | 424.38 |
| Avg. profit/mean price | 1195 | | | | |

TABLE XII.  THE RESULTS OF THE POWER SECTOR STOCKS

| Metric | JSE | NTP | PWG | SJN | TPW |
|---|---|---|---|---|---|
| Next day's pred. price | 73 | 107 | 218 | 25 | 369 |
| Huber loss | 0.00118 | 0.00076 | 0.00032 | 0.00074 | 0.00046 |
| Mean absolute error | 33.04 | 40.65 | 45.77 | 10.09 | 53.15 |
| Accuracy score | 0.9405 | 0.9393 | 0.9628 | 0.9538 | 0.9622 |
| Total buy profit | 4985 | 8098 | 14913 | 1475 | 27202 |
| Total sell profit | 4998 | 8049 | 15080 | 1470 | 28049 |
| Total profit | 9983 | 16147 | 29993 | 2945 | 55251 |
| Mean stock price | 64 | 97 | 116 | 17 | 176 |
| Total profit/ Mean price | 156 | 166 | 259 | 173 | 314 |
| Number of test cases | 538 | 791 | 645 | 520 | 688 |
| Profit per trade | 18.56 | 20.41 | 46.50 | 5.66 | 80.31 |
| Avg. profit/mean price | 214 | | | | |

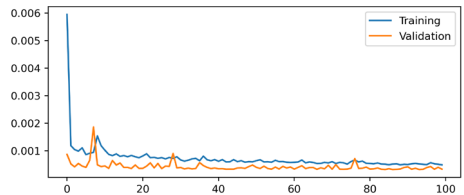

Fig. 13. The training and the validation for the stock data of Power Grid Corporation of India. *x*-axis: the no. of epochs, y-axis: the Huber loss

*Power sector:* The stocks listed in the NSE which are of critical importance in this sector are: National Thermal Power Corporation (NTP), JSW Energy (JSE), Power Grid Corporation of India (PWG), Torrent Power (TPW), and

SJVN (SJN). Table XII presents the results for the power sector stocks. Fig. 13 shows the plot of the Huber loss in the training and validation for the stock of Power Grid Corporation of India stock.

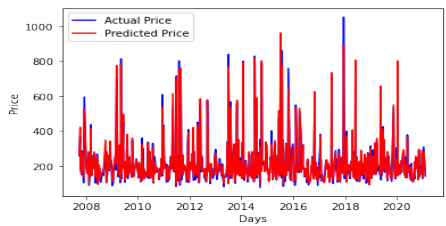

Fig. 14. The actual and the predicted *adjusted close price* of the stock of DLF over the entire training and test periods

***Realty sector:*** The stocks which are most significant in this sector and their weights in percent are: DLF-29.84, Godrej Properties (GRP)-22.28, Phoenix Mills (PHM)-12.03, Prestige Estates Projects (PRE)-7.03, and Sobha (SOB)-3.00. Table XIII presents the results for the realty sector stocks. Fig. 14 plots the actual and predicted price of the DLF stock.

TABLE XIII. THE RESULTS OF THE REALTY SECTOR STOCKS

| Metric | DLF | GRP | PHM | PRE | SOB |
|---|---|---|---|---|---|
| Next day's pred. price | 293 | 1522 | 780 | 298 | 441 |
| Huber loss | 0.00027 | 0.00029 | 0.00034 | 0.00097 | 0.00050 |
| Mean absolute error | 90.52 | 173.65 | 61.68 | 67.40 | 82.90 |
| Accuracy score | 0.9483 | 0.9628 | 0.9716 | 0.9658 | 0.9591 |
| Total buy profit | 35556 | 75922 | 67933 | 19031 | 54457 |
| Total sell profit | 35100 | 74940 | 67470 | 19592 | 54098 |
| Total profit | 70656 | 150862 | 135403 | 38623 | 108555 |
| Mean stock price | 215 | 455 | 341 | 200 | 339 |
| Total profit/ Mean price | 329 | 332 | 397 | 193 | 320 |
| Number of test cases | 658 | 537 | 668 | 497 | 684 |
| Profit per trade | 107.38 | 280.93 | 202.70 | 77.71 | 158.70 |
| Avg. profit/mean price | 314 | | | | |

TABLE XIV. THE RESULTS OF THE SMALL CAP SECTOR STOCKS

| Metric | IPC | MIN | SRF | TCM | TRN |
|---|---|---|---|---|---|
| Next day's pred. price | 1887 | 1632 | 5417 | 1061 | 786 |
| Huber loss | 0.00008 | 0.00019 | 0.00009 | 0.00037 | 0.00017 |
| Mean absolute error | 20.14 | 58.12 | 50.18 | 74.52 | 15.81 |
| Accuracy score | 0.9513 | 0.9822 | 0.9803 | 0.9671 | 0.9759 |
| Total buy profit | 264777 | 126058 | 494331 | 93402 | 70344 |
| Total sell profit | 260111 | 126038 | 486332 | 91970 | 69106 |
| Total profit | 524888 | 252096 | 980663 | 185372 | 139450 |
| Mean stock price | 330 | 406 | 800 | 370 | 145 |
| Total profit/ Mean price | 1591 | 621 | 1226 | 501 | 962 |
| Number of test cases | 1252 | 675 | 913 | 912 | 913 |
| Profit per trade | 419.24 | 373.47 | 1074.10 | 203.26 | 152.74 |
| Avg. profit/mean price | 980 | | | | |

TABLE XV. THE RESULTS OF THE TELECOM SECTOR STOCKS

| Metric | BHA | FIC | HNA | TCM | VDI |
|---|---|---|---|---|---|
| Next day's pred. price | 545 | 396 | 45809 | 1061 | 11 |
| Huber loss | 0.00039 | 0.00021 | 0.00007 | 0.00037 | 0.00054 |
| Mean absolute error | 20.75 | 19.41 | 408.26 | 74.52 | 5.63 |
| Accuracy score | 0.9441 | 0.9737 | 0.9693 | 0.9671 | 0.9718 |
| Total buy profit | 60675 | 82180 | 3686766 | 93402 | 10064 |
| Total sell profit | 59416 | 81716 | 3596337 | 91970 | 10442 |
| Total profit | 120091 | 163886 | 7283103 | 185372 | 20506 |
| Mean stock price | 274 | 160 | 6436 | 370 | 51 |
| Total profit/ Mean price | 438 | 1024 | 1132 | 501 | 402 |
| Number of test cases | 912 | 914 | 911 | 912 | 678 |
| Profit per trade | 131.68 | 179.31 | 7994.62 | 203.26 | 30.24 |
| Avg. profit/mean price | 699 | | | | |

***Small-Cap sector:*** Some of the critical small-cap stocks listed in the NSE are: Ipca Lab (IPC), Mindtree (MIN), SRF, Tata Communications (TCM), and Trent (TRN). Table XIV presents the results for the small-cap sector.

***Telecommunication sector***: Some important telecom sector stocks listed in the NSE are: Bharti Airtel (BHA), Tata Communications (TCM), Finolex Cables FIC), Honeywell Automation (HNA), and Vodafone Idea (VDI). Table XV presents the results for the stocks of the telecom sector.

***Summary***: Considering the metric average of the ratio of total profit to the mean value of the stock for the five stock as the metric for overall profitability of a sector, we observe that the FMCG sector turns out to be the most profitable sector, while the power sector is the least profitable one.

## VI. CONCLUSION

In this paper, we have proposed a deep learning regression model built of an LSTM architecture. The model is regularized using two dropout layers and has the ability to automatically scrap the wen for extracting historical stock price data using the *yahoo_fin* API. We deployed the model for analyzing 75 critical stocks from 15 sectors of the Indian stock market. An extensive set of results is analyzed on the performance of the model that demonstrates the efficacy and effectiveness of the model. The results also enable us to compare the profitability of different sectors.